\documentclass[]{emulateapj}

\begin{document}

\title{Modeling Spitzer observations of VV Ser. I. The circumstellar
disk of a UX Orionis star}

\author{Klaus M. Pontoppidan\altaffilmark{1,3,5}}
\altaffiltext{1}{California Institute of Technology, Division of Geological and Planetary Sciences, Mail Stop 150-21, Pasadena, CA 91125}
\email{pontoppi@gps.caltech.edu}

\author{Cornelis P. Dullemond\altaffilmark{2}}
\altaffiltext{2}{Max-Planck-Institut f{\"u}r Astronomie, K{\"o}nigstuhl 17, 
Heidelberg, 69117, Germany}

\author{Geoffrey A. Blake\altaffilmark{1}}

\author{A. C. Adwin Boogert\altaffilmark{1}}

\author{Ewine F. van Dishoeck\altaffilmark{3}}
\altaffiltext{3}{Leiden Observatory, P.O.Box 9513, NL-2300 RA Leiden, The Netherlands}

\author{Neal J. Evans II\altaffilmark{4}}
\altaffiltext{4}{Department of Astronomy, University of Texas at Austin, 1 University Station, C1400, 
Austin, TX 78712-0259}

\author{Jacqueline Kessler-Silacci\altaffilmark{4}}

\author{Fred Lahuis\altaffilmark{3}}
\altaffiltext{5}{Hubble Fellow}

   \begin{abstract}

We present mid-infrared Spitzer-IRS spectra of the well-known UX Orionis
star VV Ser. We combine the Spitzer data with 
interferometric and spectroscopic data from the literature covering UV to submillimeter wavelengths. The full set of data are modeled 
by a two-dimensional axisymmetric 
Monte Carlo radiative transfer code. The model is used to test the
prediction of \cite{Dullemond03} that disks around UX Orionis stars must have a self-shadowed shape, and that
these disks are seen nearly edge-on, looking just over the edge of a puffed-up
inner rim, formed roughly at the dust sublimation radius. We find that a single, relatively simple model is consistent with all the 
available observational 
constraints spanning 4 orders of magnitude in wavelength and spatial scales, providing strong
support for this interpretation of UX Orionis stars. The mid-infrared flux as measured by Spitzer-IRS is declining and 
exhibits weak silicate emission features,
consistent with a self-shadowed geometry. MIPS and SCUBA imaging shows that
the disk has a small grain dust mass as low as $0.8\times 10^{-7}\,M_{\odot}$. The low apparent dust mass may be due to
strong grain growth and settling. Further evidence for this is provided by the fact that the grains in the upper layers of the 
puffed-up inner rim must be small (0.01--0.4\,$\mu$m) to reproduce the colors ($R_V\sim 3.6$) of the extinction events, while the
shape and strength of the mid-infrared silicate emission features indicate that grains in the outer disk ($>$ 1-2 AU) are somewhat 
larger (0.3--3.0\,$\mu$m).  
From the model fit, the location of the puffed-up inner rim is estimated to be at a dust temperature of 1500\,K or at 
0.7--0.8\,AU for small grains. This is almost twice the rim radius estimated from
near-infrared interferometry. Since larger (more grey) grains are able to penetrate closer to the star for the same
dust sublimation temperature, a 
plausible interpretation of the data is that these larger grains have settled to the disk mid-plane in the puffed-up inner rim. 
A best fitting model for the inner rim in which large grains in the  
disk mid-plane reach to within 0.25\,AU of the star, while small grains in the disk surface create a puffed-up inner
rim at $\sim 0.7-0.8$\,AU, is able to reproduce all the data, including the near-infrared visibilities. 
   \end{abstract}


\section{Introduction}
Edge-on and nearly edge-on disks around Herbig Ae and T Tauri stars are
ideal objects for studying the structure and composition of protoplanetary
disks. In addition to infrared emission from the disk, one has supplemental
information from absorption of starlight along the line of sight through the
surface layers or the interior of the disk. This probing technique has
been used by various authors to analyze the dust size distribution
in such disks \citep{Wood02,Duchene03,Wolf03,Watson04} and to determine the abundance of icy grain in the
outer regions of these disks \citep{Thi02, Pontoppidan_crbr}. 

Because of the additional constraints compared with other disks, it is clear that edge-on disks are
ideal targets for detailed studies involving spectroscopy and images at many
wavelengths. Given the many new sensitive observational facilities, including space-based
mid-infrared imaging and spectroscopy as well as near-infrared interferometry, a powerful approach to analyzing these data is the simultaneous fitting with a multi-dimensional continuum radiative transfer model of
the disk. 

A particularly interesting sub-class of (presumably) nearly edge-on disks is ``UX Orionis
stars''. For these objects the stellar flux on average is only
marginally extincted, but they undergo frequent, highly non-periodic extinction
events, lasting a few days to weeks, in which the star dims strongly and
becomes redder. A model in which dusty clumps in Keplerian orbits 
temporarily obscure the central star, as originally proposed by \cite{Grinin88}, seems 
the best explanation for the observational constraints. In particular, in the bottom of very
deep minima, the color track eventually turns around and the colors become bluer. In the context
of the ``dust clump'' model, this characteristic ``blueing'' in deep minima signifies that the stellar photosphere 
has disappeared entirely from view and
that scattered light from the disk surface begins to dominate in the optical wavebands. 

While it has been suggested that these extinction events may be due to comets
passing in front of the star \citep{Grady97}, 
others have argued for a
nearly edge-on disk scenario in which one narrowly looks over the surface of
the disk toward the star, where small hydrodynamic perturbations
temporarily pass through the line of sight \citep{Bertout00, Natta00}. 
The short time scales for these extinction events indicate
that the perturbations should happen close to the dust evaporation radius. It was argued by \cite{Natta01} and \cite{Dullemond03} 
that the puffed-up inner rim of the dusty part of the disk
\citep[see also][]{Dullemond01} is likely to be the location where
these extinction events are produced. The inclination of the disk is such that
the line of sight passes just over (or just through) the upper edge of the
puffed-up inner rim, and hydrodynamic fluctuations can then vary the
extinction of starlight along the line of sight on a time scale consistent
with what is observed. Since the central star is typically not strongly attenuated outside a UX Orionis event, 
this scenario works only if the line of sight does
not pass through many magnitudes of extinction in the outer regions of the disk. Hence, the
outer parts of the disk cannot have a flaring geometry for these sources.

The conclusion reached by \cite{Dullemond03} was that all UX Orionis stars should be self-shadowed disks, in which
the outer disk is geometrically thin enough to lie entirely in the shadow
cast by the puffed-up inner rim, and therefore does not intersect a line of
sight passing just over the edge of this rim. A marginal flaring, in
which the outer disk just barely appears above the inner-rim shadow, may
be possible as well, as long as the resulting extinction along
the line of sight is marginal. It was shown by these authors
that the shape of the spectral energy distributions of all known UX Orionis
objects indeed indicate that they are mostly non-flared (self-shadowed)
disks, with some exceptions lying just on the border between flaring and
self-shadowing disks. This gives strong support for the ``inclined disk'' scenario.

As part of the Spitzer Legacy program {\it `From Molecular Cores to Protoplanetary Disks'} (c2d) \citep{Evans03},
we obtained 5.2--37.0\,$\mu$m spectroscopy as well as IRAC and MIPS images of 
the well-known UX Orionis star VV Ser. The star was originally singled out
for further study due to the presence of an extended (over $4\arcmin$) mid-infrared nebulosity surrounding it. 
This nebulosity is analyzed in detail in a companion paper \citep[][hereafter Paper II]{paperII}, in
which it is argued that the nebulosity is due to the quantum heating of Polycyclic
Aromatic Hydrocarbons (PAHs) as well as very small silicate or carbon grains. Imprinted
on the nebulosity is a wedge-like dark band which we interpret as a shadow
cast by a small (less than a few 100 AU) nearly edge-on disk \citep[see also][]{Pontoppidan_shadow}. The presence of a disk shadow
allows a relatively accurate determination of the inclination and position angle of the system, which is crucial
for the interpretation of VV Ser as a UX Orionis star.

In this paper we concentrate on analyzing VV Ser in terms of the UX Orionis phenomenon. 
To do this we create an axisymmetric radiative transfer model using a wide range of data to constrain the
disk structure. This includes not only the {\it Spitzer} data, but also 
JCMT-SCUBA submillimeter imaging, near-infrared interferometric 
visibilities and 
optical light curves from the literature. 
Although some aspects of an axisymmetric model cannot be uniquely constrained, 
we can make firm conclusions about the disk geometry, grain
sizes, inner rim structure, etc. More importantly, our analysis gives
further strong evidence in favor of the scenario for UX Orionis stars put
forward by \cite{Dullemond03}. In doing so, this lends
support for the puffed-up inner rim model of \cite{Natta01} and
\cite{Dullemond01}, as well as for the interpretation of the
near-IR to far-IR slope of the SED in terms of flaring versus self-shadowed
disks \citep{Dullemond04, Meeus01}. This scenario allows us to interpret all the observed data
of VV Ser in terms of a {\em single} disk model with a
simple geometrical interpretation.

We construct an axisymmetric radiative transfer model of the
entire system from 0.5 to 50,000\,AU scales. The main objectives
are to test the scenario in which the UX Ori phenomenon is caused by a nearly
edge-on disk, whether a disk shadow can plausibly be projected into the
mid-infrared PAH nebulosity (see Paper II) and finally to use the model to constrain the
structure and dynamics of the puffed-up inner rim. 

This method necessarily targets specific objects for detailed modeling, and one 
should take care when generalizing from one specific object to a class of disks.
One central point to this study is to determine observable predictions that can be 
applied to other UX Orionis stars, or even proto-planetary disks in general. 

In this paper we first describe the
observations that have been collected (Sect. \ref{observations}) and the characteristics of the source 
(Sect. \ref{source} ). In Sect. \ref{model} the radiative transfer 
model is presented. Sect. \ref{explained}
discusses how each observational constraint is implemented into the model as well as possible sources of degeneracies. 

\section{Observations}
\label{observations}

The primary constraints on the model of VV Ser 
are provided by mid-infrared spectroscopy and imaging
obtained with the {\it Spitzer Space Telescope} \citep{Werner04}. Mid-infrared spectra of VV Ser were 
obtained with the Spitzer Infrared Spectrometer, IRS \citep{Houck04} using the Short-Low (SL) module from 5.2--14.5\,$\mu$m, the Short-High 
module from 9.9--19.6\,$\mu$m and the Long-High module from 19--37\,$\mu$m. The
spectra were reduced using the pipeline version S11.0.2 and extracted with the c2d extraction routines. 
The SL spectra were extracted in a 12 pixel aperture. The background was estimated using the entire length
of the slit and subtracted using a high-order polynomial fit to the measured background in order to minimize 
any additional noise from the background subtraction. The SH and LH spectra were extracted using full aperture
extractions. The short slits in the high resolution modules cause any background subtraction to be highly
PSF-model dependent. Consequently, the background was not subtracted from the SH and LH spectra. Since VV Ser
is a bright source between 10 and 40\,$\mu$m, we estimate the background to be a minor contribution, 
even considering the extended emission
present around the source (see Paper II). The orders and modules were matched by scaling to the shortest wavelength order.
The scaling factors applied were 5-10\%, and the absolute flux level of the spectrum is 
therefore considered accurate only to 10-20\%. 
The spectrum of VV Ser has AOR Key 0005651200.
We present IRAC and MIPS imaging at 3.6, 4.5, 5.6, 8.0, 24 
and 70\,$\mu$m of the area surrounding VV Ser in Paper II. 

In support of the Spitzer observations, additional continuum data at 850 and 450~$\mu$m were obtained in service mode
with the Submillimetre Common User Bolometer Array (SCUBA) on the
James Clerk Maxwell Telescope (JCMT)\footnote{The JCMT is operated by the
  Joint Astronomy Centre in Hilo, Hawaii on behalf of the parent
  organizations: the Particle Physics and Astronomy Research Council
  in the United Kingdom, the National Research Council of Canada and
  the Netherlands Organization for Scientific Research.} on 2005
January 27. VV Ser was observed in a 64 point jiggle map
(2.3\arcmin\ field) with approximately one hour of integration. The
weather was good with 225~Gz sky opacities of about 0.05. The
pointing was checked regularly and found to be accurate to within a
few arcseconds. Maps of CRL2688 and Mars were used for calibration
purposes. The absolute calibration is accurate to about $\pm
20$\%. The beam size (HPBW) of the SCUBA observations is approximately 14$''$
at 850~$\mu$m. No continuum emission was detected toward VV Ser
with a 3$\sigma$ upper limit of 10~mJy~beam$^{-1}$.

An ultraviolet spectrum obtained with the International 
Ultraviolet Explorer (IUE) was taken from the Multimission Archive at STScI (MAST)\footnote{
Some of the data presented in this paper were obtained from the 
Multimission Archive at the Space Telescope Science Institute (MAST). 
STScI is operated by the Association of Universities for Research in Astronomy, Inc., 
under NASA contract NAS5-26555. Support for MAST for non-HST data is provided by the 
NASA Office of Space Science via grant NAG5-7584 and by other grants and contracts.}. 
More than 12 years of photometric monitoring in the optical UBVRI bands was taken from the 
catalogue of \cite{Rostopchina01}. The photometry for the VV Ser point source is summarized in
Table \ref{photometry}.

\begin{table}
\centering
\caption{Photometry of VV Ser}
\begin{tabular}{lll}
\hline
\hline
Wavelength ($\mu$m)&Flux [Jy]$^1$&Reference\\
\hline
0.2&$(4\pm 2) \times 10^{-4}$&IUE\\
0.365&$0.011\pm0.0005$&\citep{Rostopchina01}\\
0.44&$0.035\pm0.002$&\citep{Rostopchina01}\\
0.55&$0.077\pm0.004$&\citep{Rostopchina01}\\
0.70&$0.14\pm0.007$&\citep{Rostopchina01}\\
0.90&$0.28\pm0.014$&\citep{Rostopchina01}\\
1.235&$0.47\pm0.02$&2MASS\\
1.662&$0.97\pm0.05$&2MASS\\
2.159&$1.87\pm0.09$&2MASS\\
3.79&$3.0\pm0.1$&\citep{Berilli92}\\
4.64&$3.4\pm0.2$&\citep{Berilli92}\\
5.5&$4.2\pm0.1$&Spitzer-IRS, this paper\\
8.0&$4.5\pm0.1$&Spitzer-IRS, this paper\\
12.0&$4.2\pm 0.2$&Spitzer-IRS, this paper\\
25.0&$3.4\pm 0.2$&Spitzer-IRS, this paper\\
36.0&$2.7\pm 0.2$&Spitzer-IRS, this paper\\
70.0&$0.35\pm 0.1$&Spitzer-MIPS, paper II\\
450&$<0.18 (3\sigma)$&JCMT-SCUBA, this paper\\
850&$<0.03 (3\sigma)$&JCMT-SCUBA, this paper\\
\hline
\end{tabular}
\begin{itemize}
\item[$^1$] The optical photometry is for the ``quiescent'' state.
\end{itemize}
\label{photometry}
\end{table}

\section{Source description}
\label{source}

VV Ser is a typical UX Orionis star located in the Serpens molecular cloud. The adopted distance
of VV Ser is $\sim 260\,$pc as determined for the Serpens molecular complex by \cite{Straizys96}. There is 
some variation in the literature regarding the effective temperature of VV Ser, although most
classify the star as a late B-type. \cite{Hernandez04} identify VV Ser as a B6 star due to the presence
of a number of He lines in their low-resolution spectra. \cite{Mora01} find a spectral type of A0 using high 
resolution spectroscopy.  
The exact choice does have measurable consequences for the model SED at wavelengths below $\sim 1\,\mu$m. 
The optical colors favor a star closer to A0 than B6. The stellar luminosity of 49\,$L_{\odot}$ also suggests 
that VV Ser is of later type since a B6 classification puts it below the ZAMS in the HR diagram \citep{Siess00}.

Since the optical-UV spectrum of VV Ser does not appear to be strongly veiled and is of high quality, it is possible
to derive an accurate stellar luminosity of $(49\pm5)\times (d/260\,\mathrm{pc})^2\,L_{\odot}$. 
Using the evolutionary tracks of \cite{Siess00} and an effective temperature of 10,200\,K, this gives
a stellar mass of $2.6\pm0.2 \,M_{\odot}$ and an age of $3.5\pm0.5 $\,Myr for a solar metallicity. The uncertainties
reflect the uncertainties in the luminosity and $T_{\rm eff}$ rather than those of the model tracks. 

The optical colors of the star correspond to a steady extinction of $A_V \sim 3$, but with frequent, 
non-periodical dips lasting of order 10 days with brightness minima corresponding to 
0.5--4.0 mag of additional extinction. 

\section{Model}
\label{model}
To model the observed SED and Spitzer imaging of VV Ser, we use the
axisymmetric Monte Carlo radiative transfer code RADMC \citep{Dullemond04}
in combination with the raytracer of the more general code RADICAL 
\citep{Dullemond00}. 
The density structure is axisymmetric, but the photons are followed in all three dimensions.  
The code is used to derive the temperature structure of a given dust 
distribution. The dust temperature is determined for a passive disk, i.e. it is assumed that
accretion heating is negligible relative to direct stellar irradiation. 
This is justified by the high luminosity of the central star (49\,$L_{\odot}$), since the accretion luminosity is $\lesssim 3\,L_{\odot}$
for accretion rates $\dot{M}\lesssim 10^{-7}\,M_{\odot}\,\rm year^{-1}$ \citep{Kenyon93}. 
The low mass of the disk [$8\times 10^{-6}\,M_{\odot}\times (R_{\rm gas to dust}/100)$], where $R_{\rm gas to dust}$ is the
gas to dust mass ratio (see. Sect. \ref{constraining}) also limits the possible range of accretion rates, since only a small fraction of the
entire disk mass can be expected to accrete per year. The low mass of the disk is particularly well constrained by the 70\,$\mu$m
MIPS point as well as the upper limit to the 850\,$\mu$m flux. 

Once a temperature structure has been determined, RADICAL can be used to
create SEDs and images. The code is restricted to isotropic scattering. Taking
non-isotropic scattering into account may change the optical colors of the model somewhat, but
without resolved optical images of the disk, non-isotropic scattering effects cannot be properly constrained.   
One advantage of the code is the ability to include an arbitrary number of dust components, each
with a unique opacity.
The setup has been used to model similar protostellar disks
\citep{Pontoppidan_crbr, Pontoppidan_shadow}, while the code itself has been extensively tested for 
other applications \citep{vanBemmel03,Dullemond05,Meijerink05} and has recently been benchmarked relative to independent
continuum radiative transfer codes \citep{Pascucci04}. 

\subsection{Disk geometry}
VV Ser is modeled using a density structure consisting of a central
disk surrounded by a spherically symmetric envelope. For the central disk, the
following density structure is adopted:

\begin{equation}
\rho(R,Z) = \frac{\Sigma(R)}{H_p(R)\sqrt{2\pi}}\exp\left(-\frac{Z^2}{2H_p(R)^2}\right),
\label{diskeq}
\end{equation}
where $\Sigma(R)=\Sigma_{\rm disk}\times(R/R_{\rm disk})^{-p}$ is the 
surface density and 

\begin{equation}
H_p(R)/R=(H_{\rm disk}/R_{\rm disk})\times(R/R_{\rm disk})^{\alpha_{\rm fl}} 
\label{scaleeq}
\end{equation}
is the disk scale height with $\alpha_{\rm fl}$
the flaring index. The flaring index is $2/7$ for a passive irradiated disk with a grey dust opacity \citep{CG}.
For $\alpha_{\rm fl}\lesssim 0$ the disk is non-flaring and shadowed. We have chosen an intermediate value of $1/7$. However, once the disk
is self-shadowed, the exact value of the flaring parameter does not have a strong influence on the
SED. Eq. \ref{diskeq} is appropriate for the
vertical structure of an isothermal disk in hydrostatic equilibrium
and gives a convinient parametrization, even for disks that are non
isothermal and not in hydrostatic equilibrium.

Indeed, we do not require a vertical structure determined by hydrostatic equilibrium such as that described in 
\cite{CG} and \cite{Dullemond01}, but let the outer vertical scale height be a free parameter. 
While the unique solution of \cite{CG} assumes that gas and dust are well-mixed and thermally coupled, there are
several mechanisms that may operate to cause the vertical structure of the dust to deviate significantly from
hydrostatic equilibrium.  
For instance, dust grains may decouple from the gas by grain growth and settling in which
case hydrostatic equilibrium no longer applies. Recent modeling of dust settling indicates that it is a rapid process
in circumstellar disks and produces highly observable effects \citep{Dullemond04}. 

For the structure of the inner disk, we adopt a puffed-up inner rim model similar to that 
of \cite{Dullemond01}. The scale height of the inner rim, $H_{\rm rim}$, is included as an adjustable parameter. The radius, 
$R_{\rm rim}$,
of the inner rim is set to the location at which a temperature of 1500\,K is reached. Note that since the
dust in the inner rim does not have a grey opacity, the radius of the inner rim is pushed to a significantly
larger radius than that of grey dust. The puffed-up inner rim is connected to the ``normal'' disk structure
by changing the scale height power law of Eq. (\ref{scaleeq}) at a few times the inner rim radius such that the scale height
rises inwards to meet the puffed-up inner rim. The radius of the break in the power law is not 
strongly constrained, and a presumably reasonable value of $2.5\times R_{\rm in}$ is adopted here. 

The disk is surrounded by a large ``envelope'' simulating the presence of the large-scale molecular cloud.
The column density of the cloud is adjusted to reproduce the measured extinction toward VV Ser outside
of extinction events. A cavity centered on the star+disk system is carved out of the cloud.
Paper II discusses the model fit on large spatial scales, including the detailed structure of the envelope, and how a local
model grid for the envelope parameters was calculated, while this paper deals with the parameters of the disk itself. 

\subsection{Dust opacities}
\label{dustopacities}
The mid-infrared spectrum and the optical colors indicate that  
several different dust opacities must be included in the 
model, corresponding to different parts of the disk. The silicate emission band at 9.7\,$\mu$m is broader and flatter than
an interstellar feature (see Sect. \ref{midir}), indicative of grain growth \citep{Bouwman01,Kessler06}, 
whereas the colors of
the flux variations at optical wavelengths indicate that the grains in the innermost
parts of the disk are not very different from small interstellar grains. Grains large enough to produce a flat silicate feature will
have grey opacities in the optical wavebands.
Finally, the PAHs necessary to model the large nebulosity surrounding VV 
Ser contribute an opacity of their own (see paper II for details). 

The model allows for an arbitrary number of distinct dust reservoirs with relative abundances that
may vary throughout the grid. In the VV Ser disk, we use two distinct dust components: The first consists
of small grains that produce non-grey opacities in the optical wavebands, the second component consists of larger $\sim 1\mu$m grains. 
The small-grain dust is located in the puffed-up rim ($\lesssim 1\,$AU), while the rest of the disk is populated with the larger 
grain component. 

The opacities of the thermalized dust grains are calculated using Mie theory. The grains are spherical
silicates with inclusions of carbonaceous material. The silicate optical constants are those
of oxygen-rich silicates from \cite{OH94}, while the carbon optical constants are of carbon clusters formed at 800\,K from \cite{Jaeger98}.
We do not include any ice component in the dust opacity.  
The Maxwell-Garnett 
effective medium formula \citep[e.g.][]{BH} has been used to calculate the optical constants of 
the silicate with carbon inclusions. The volume fraction
of the carbon is 30\%, roughly consistent with that determined by 
\cite{Draine03}. The final opacities are then calculated with a grain size
distribution consisting of a power law: $dN(a)/da\propto a^{-3.5}$
with minimal ($a_{\rm min}$) and maximal 
($a_{\rm max}$) grain sizes. The disk SED is attenuated by foreground dust from an extended envelope, which
is discussed in detail in Paper II. 
The exact choices of grain size distributions for the disk dust are justified in Sects. \ref{optical} (optical colors) 
and \ref{midir} (silicate emission feature), and the dust parameters are summarized in Table \ref{dustcomp}.
For comparison to other dust opacities, 
it is useful to note that the ``large grain'' dust opacity of the disk has 
$\kappa_{\rm 850\,\mu m}=0.01\,{\rm cm^{2}\,g^{-1}}\times (R_{\rm gas to dust}/100)$, where
$R_{\rm gas to dust}$ is the gas to dust ratio.
This is similar to, but on the small side of the opacities (for coagulated grains) of \cite{OH94}. 

\section{Observed features explained by the model}
\label{explained}

\subsection{Constraining the model}
\label{constraining}

\begin{figure}
  \includegraphics[width=9cm]{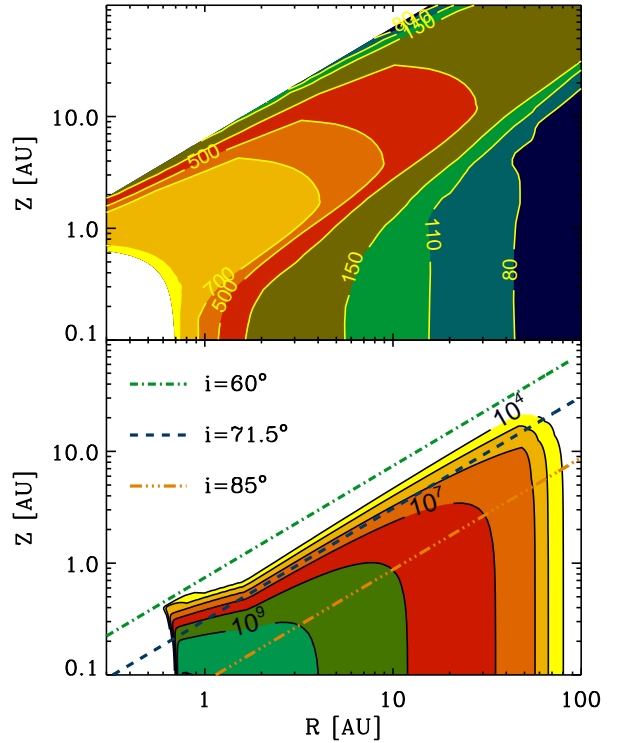}
  \caption{Physical structure of the disk model. {\it Top panel:} The temperature of the disk as 
calculated by RADMC in Kelvin. {\it Bottom panel:} The density structure of the disk in units of hydrogen molecules
per cm$^3$. The three
lines indicate the path from the central star to an observer for different inclination angles. Note that for the best-fitting inclination angle
of 71.5$\degr$, the line of sight to the central star passes through both the puffed-up inner rim as well as part of the outer disk. }
  \label{diskstruct}
\end{figure}

Due to the long computing times for a single model, 
the model parameters are varied by hand. Rather than calculating a comprehensive grid in all
parameters, hundreds of models were calculated until a good fit was found to all the observable
quantities. While this is not an optimal
method for quantifying any degeneracies in the fit, 
some parameters may be independently constrained on physical grounds. 
In Table \ref{parameters}, the parameters of the best-fitting model are summarized. This
model has been constructed using all available observational constraints apart from
the near-infrared interferometry visibilities discussed in Sect. \ref{interferometry}. We will refer to this model
as ``Model 1''. 
The density and temperature structure of Model 1 are shown in Fig. \ref{diskstruct}. The SED of Model 1 is
compared to the observed SED of VV Ser in Fig. \ref{SED}. 
In Sect. \ref{extinction_events}--\ref{midir}, we discuss how each observation property is connected to the model. In Sect. \ref{interferometry}, it is discussed
how the near-infrared interferometry from the literature requires Model 1 to be modified on small scales.

\subsection{Extinction events}
\label{extinction_events}
\begin{figure*}
  \centering
  \includegraphics[width=5cm,angle=90]{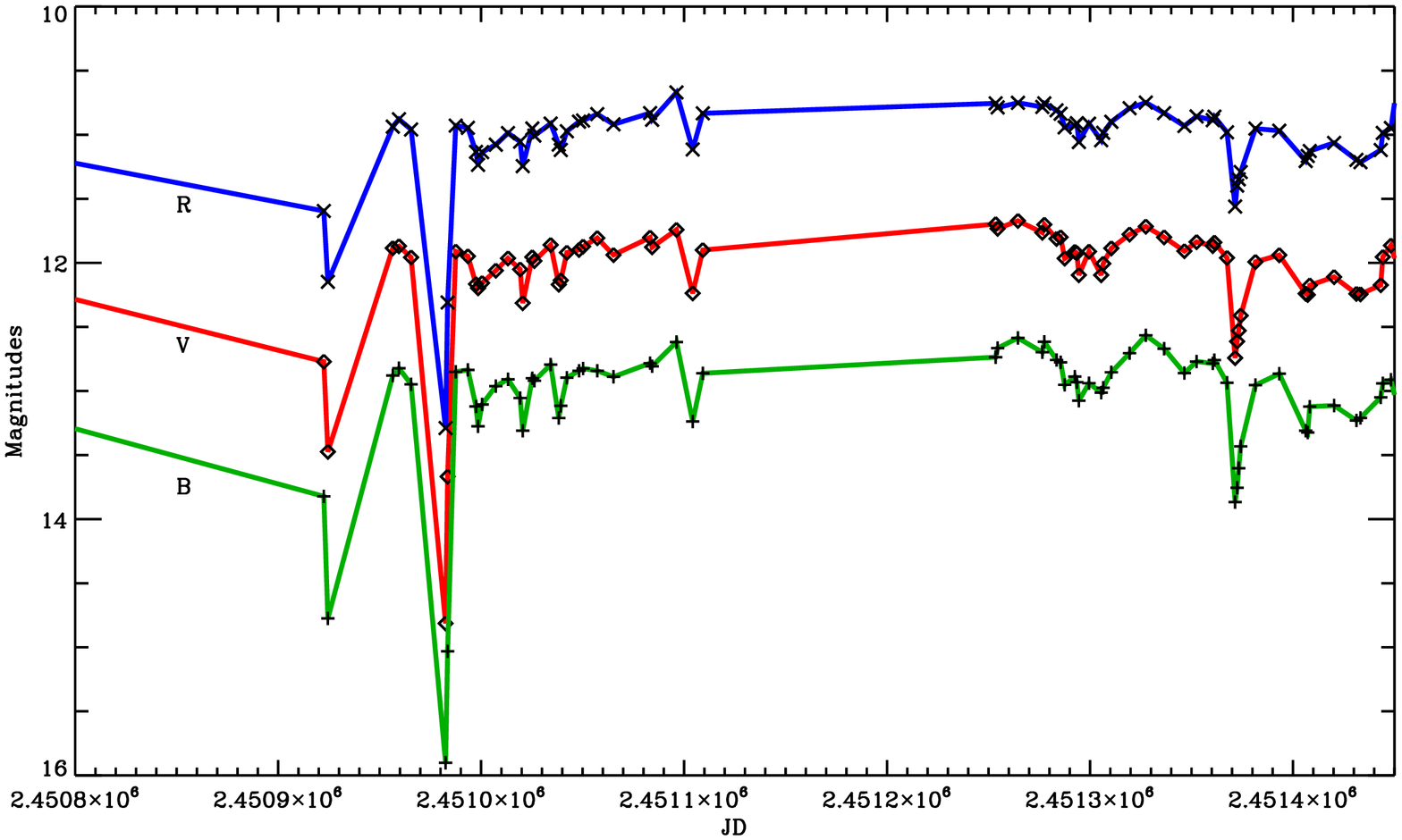}
  \includegraphics[width=7cm]{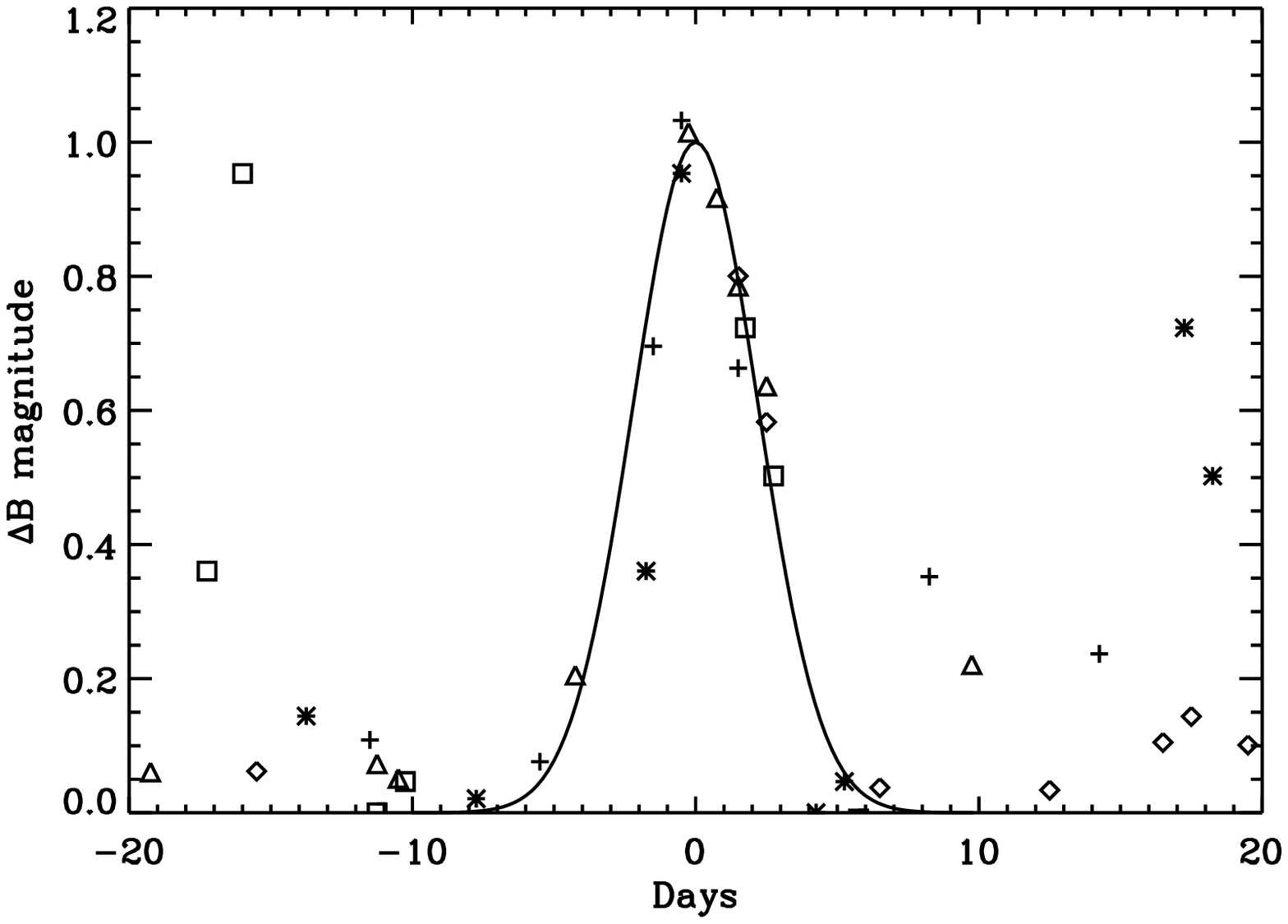}

  \caption{{\it Left panel:} The optical light curve of VV Ser from the catalogue of \cite{Rostopchina01} spanning 
the best-sampled 2 years. {\it Right panel:} The average profile of the extinction events over the best-sampled 2 year period. The curve is a Gaussian
with a FWHM of 5.5 days. }
  \label{lightcurve}
\end{figure*}

The optical light curve of VV Ser, observed by \cite{Rostopchina01}, is shown in Fig. \ref{lightcurve}. It is seen that
the light curve of VV Ser contains a number of extinction events, each lasting a few days. Although
each event is undersampled, their average duration can be estimated
by searching for the duration most compatible with all the observed 
events. Assuming that the extinction events have an identical 
Gaussian shape in the light curve and vary only in intensity, it is found that 
the events are consistently shorter than 5.5 days and longer than 3.0 days 
(FWHM). The best-fitting Gaussian is shown in the right panel of Fig. \ref{lightcurve}. 
The width of any perturbation of the inner rim causing an
extinction event is likely to be less than or comparable to 
the pressure scale height of the disk. Scenarios where this is not the case are 
interesting to consider, but probably require hydrodynamical simulations to constrain. 
Assuming Keplerian rotation, this requirement constrains the radius
of the inner rim of the disk, since a rim radius that is too small will cause extinction 
events to be shorter than those observed. An inner
radius corresponding to a dust sublimation temperature of 1500\,K for small grains is located
at 0.7--0.8\,AU at the VV Ser luminosity of $\sim 49$\,$L_{\odot}$. 
This corresponds to a rotation period of the inner rim of $\sim 180$ days and
a physical size of the dust perturbation causing the extinction in the azimuthal 
direction of 0.09--0.15\,AU. A spherical dust perturbation then has an average density
of 2--3\,$\times 10^9$\,cm$^{-3}$ in order to produce a maximum observed
extinction of 5 magnitudes in the $V$ band. These values are close to those of
the model with an inner rim scale height of  $H_{\rm rim}/R_{\rm rim}=0.105$ 
and a maximum gas density along the line of sight of $\sim 10^9$\,cm$^{-3}$ between extinction events for an inclination
of $\sim 70\degr$. A specific extinction event then 
corresponds to creating an over-density in the inner rim enhanced by a factor of a few compared to
the ``quiescent'' state of the inner disk. 

\subsection{The outer radius}
The outer radius of the disk is not well constrained and is somewhat degenerate with the disk opening angle.
The mass is constrained by the 70\,$\mu$m MIPS point and the 850\,$\mu$m upper limit to be as low as 
$8\times 10^{-6}\times (R_{\rm gas to dust}/100)\,M_{\odot}$. 
As usual, it must be stressed that the observations do not probe grains that have grown to sizes of 
more than a few mm, and the disk may contain a significant population of such grains in the mid-plane. 
Similarly, any total mass estimate will necessarily assume a dust-to-gas ratio of 100, but 
this is likely to be very different for a disk as evolved as VV Ser. The mass of small dust grains is accurate to
the uncertainty in the far-infrared opacity, i.e. by a factor of a few.

\begin{table}
\centering
\caption{Adopted dust components}
\begin{tabular}{llll}
\hline
\hline
Component&$a_{\rm min}$ [$\mu$m]&$a_{\rm max}$ [$\mu$m]&Disk regime\\
\hline
\multicolumn{4}{l}{Model 1 (Standard model) and Model 2}\\
\hline
Small grains&0.005&0.4&$R<1\,$AU, $R>5000\,$AU\\
Large grains&0.3    &3.0   &$5000>R>1\,$AU\\
\hline
\multicolumn{4}{l}{Model 3}\\
\hline
Small grains&0.005&0.4&$0.7<R<1\,$AU, $R>5000\,$AU\\
Large grains&0.3    &3.0   &$R<0.7$, $5000>R>1\,$AU\\
\hline
\end{tabular}
\label{dustcomp}
\end{table}

\begin{table*}
\centering
\caption{Best-fitting model parameters}
\begin{tabular}{lllll}
\hline
\hline
Parameter & Model 1 & Model 2 & Model 3 & Estimated range (see text)\\
\hline
Lum.&49\,$M_{\odot}$&--&--& 45--55\\
Sp. T.&B9 (10\,200\,K)&--&--&A2--B6\\
Dust mass&$0.8\times 10^{-7}\,M_{\odot}$&--&--& 0.1--8.0$\times 10^{-7}\,M_{\odot}$\\
$p$&$-1$&--&--&not explored\\
$\alpha_{fl}$&1/7&--&--&not explored\\
$H_{\rm disk}/R_{\rm disk}$&0.125&--&--&0.1--0.15\\
$H_{\rm rim}/R_{\rm rim}$&0.105&--&--&0.1--0.15\\
$T_{\rm sub}$&1500\,K&2000&1500&1300--1600\,K\\
$R_{\rm rim}$&0.7\,\rm AU&0.4&0.27 and 0.80&given by $T_{\rm sub}$\\
$R_{\rm disk}$&50\,AU&--&--&20--200\,AU\\
Incl. & $71.5\degr$&$40\degr$&$68\degr$&65--75\degr\\
P.A. &$11\degr$ &$170\degr$&$15\degr$&10--20\degr (see Paper II)\\
\hline
\end{tabular}
\label{parameters}
\end{table*}

\begin{figure}
  \plotone{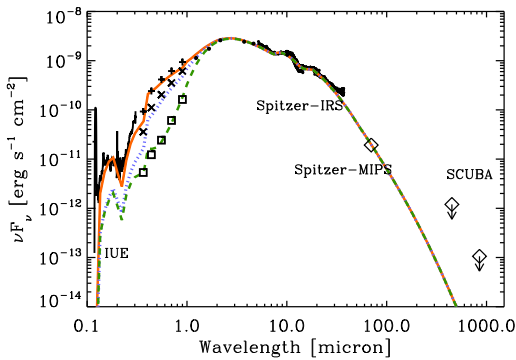}
  \caption{The observed SED of VV Ser compared to SEDs of Model 1. The three
model SED curves correspond to a varying optical depth of the perturbation of
the inner rim responsible for the observed UX Ori extinction events (the solid curve corresponds to $\Delta A_V=0$, the dotted curve
to $\Delta A_V=0.75$ and the dashed curve to $\Delta A_V=4.0\,$mag.). The sets of optical data corresponding
to the symbols (+), (x) and ($\square$) are single points in the light curve obtained at JD2450624.358, JD2450683.290 and JD2450956.499, respectively. }
  \label{SED}
\end{figure}

\begin{figure*}
\centering
  \plotone{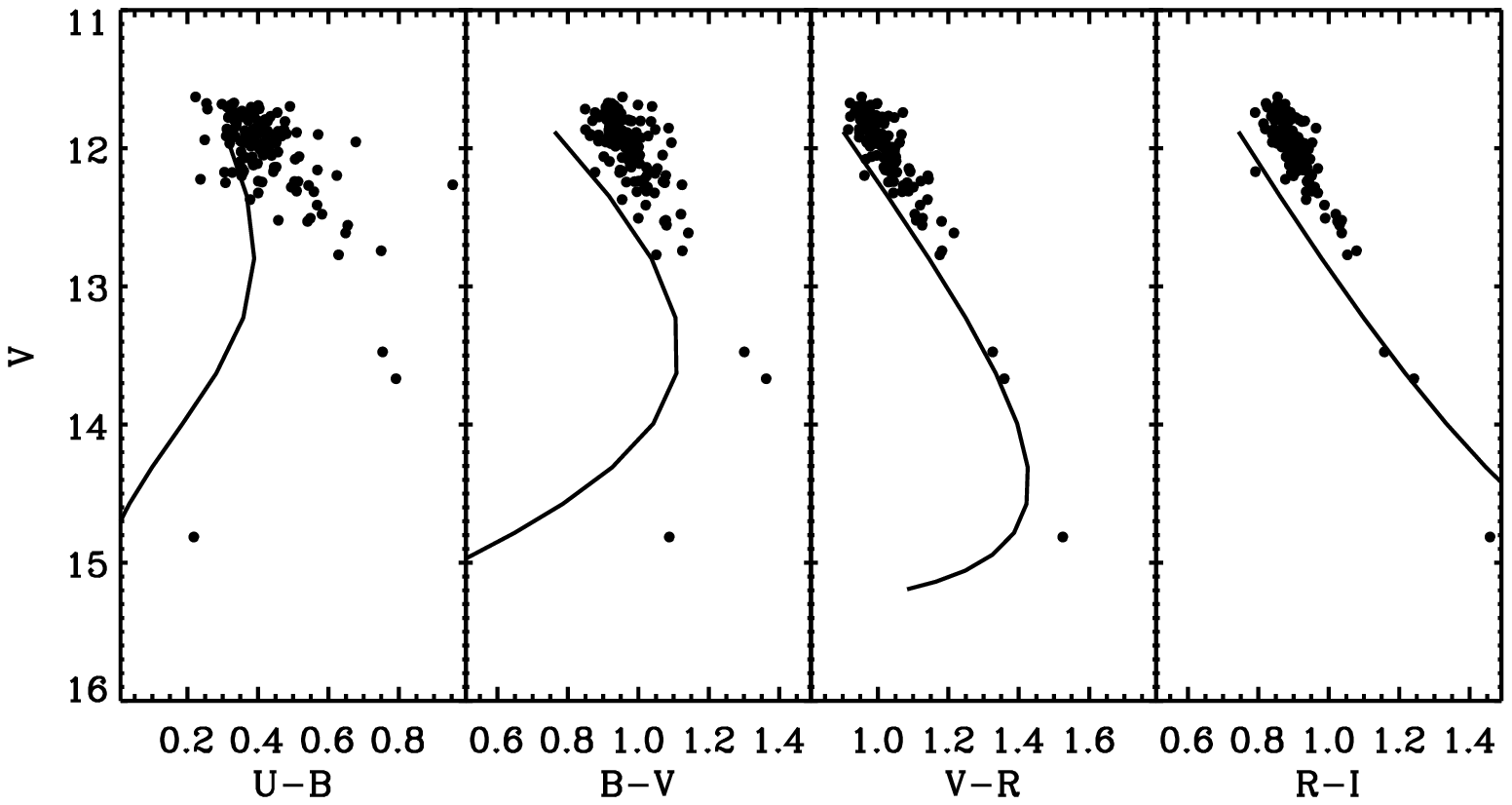}
  \caption{Color-magnitude diagrams of the photometry of VV Ser from \cite{Rostopchina01} compared to
curves calculated using the radiative transfer model. The model curves correspond to extinction events of $A_V=0-4$\,mag. 
The small 0.15 mag offset in the R-I color is due to an inaccuracy in either to model photospheric color
or the inner rim color.}
  \label{colormag}
\end{figure*}

\subsection{Optical photometry and variability}
\label{optical}
As discussed in Sect. \ref{dustopacities}, the observation that the grains 
responsible for the silicate emission features appear to be different from those 
responsible for the extinction events is taken into account in the model.
While many UX Orionis stars appear to have disks in which the dust causing the extinction events is dominated by 
small grains ($a\lesssim1\,\mu$m) \citep{vandenancker99}, models of grain growth predict that
the inner parts of disks very quickly become dominated by large grains \citep{Dullemond05}.
A priori, it is not clear why the grains responsible for the extinction events have a 
different size distribution than those of the outer disk. As a first attempt, we simply model the observations
by changing the opacity of the dust in the puffed-up inner rim at radii between the rim and
the break in the scale height power law defined in Sect. \ref{model} (see Table \ref{dustcomp}). 
This has the effect of converting the grey (in the UV--NIR) disk opacity to one reproducing the optical colors.
Note that this also causes the inner rim to absorb visible photons more efficiently than it
re-emits the energy in the near-infrared.
This significantly increases the temperature of the star-facing surface of the rim and has the effect of pushing
the rim outwards, given a certain dust sublimation temperature. Possible alternative geometries are discussed in
Sect. \ref{interferometry}. 

The grain model of the material causing the extinction events is constrained 
by the optical extinction curve that can be derived by the observed optical 
reddening vectors. The extinction events correspond to grains that appear to
have properties similar to those of interstellar grains. 
The reddening vectors in the optical 
color-magnitude diagrams are fitted by a power-law grain size distribution by
varying the minimum and maximum sizes as well as the power law index. 
Not unexpectedly, a good fit is found for $a_{\rm min}=0.05\,\mu$m, 
$a_{\rm max}=0.4\,\mu$m, giving an optical extinction law with $R_V\sim 3.6$.

As discussed in Sect. \ref{constraining} the perturbations of the inner rim are presumably not
wider in the azimuthal direction than the pressure scale height. In essence
this means that a dust perturbation in a Keplerian orbit roughly subtending $10\degr$ as seen from
the star is responsible for an extinction event. The optical variability of  
VV Ser is therefore fundamentally a three-dimensional effect and to model it within the
axisymmetric framework presented here requires some assumptions. If it is assumed
that the azimuthal angle the dust perturbation subtends as seen from the star is $\ll 2\pi$, 
and that the shadowing and scattering effects caused by the perturbation will not affect the thermal part of the SED. 
In this case, only the extinction at a specific inclination through the perturbation needs to be taken into account.
This means that the temperature structure of the system can be calculated for the ``quiescent'' 
state of the inner rim in which no localized perturbation of dust is providing extra extinction toward the
star. The ``active'' state corresponding to an extinction event is simulated by using the
temperature structure calculated by the Monte Carlo code, but adding extra extinction to the input
spectrum of the central star when calculating
the final SED by raytracing. Thus, by attenuating the star by a variable amount of dust,
model tracks through the optical color-magnitude diagrams can be constructed
and compared to the observed tracks. The effect on the SED of attenuating the input stellar spectrum by a variable 
extinction ($\Delta A_V$) is shown in Fig. \ref{SED} and compared to photometric data obtained during and outside of extinction events.
 
In the model for VV Ser, a blueing effect occurs naturally because scattering dominates the
optical flux for high optical depths through the extincting blob of dust. The resulting model 
tracks are plotted on the observed photometric points for VV Ser in Fig. \ref{colormag}. 
The tracks correspond to $A_V$'s of 0.0--4.0\ mag.
It is seen that a strong blueing effect appears for $A_V\gtrsim 2.5$\ mag. The blueing is strongest
in the $B-V$ color. In $R-I$ a slight reddening is seen instead. This is due to the 
$I$ band photons being slightly dominated by thermal emission from the 1500\ K inner 
rim rather than from scattering of photospheric photons. The model approximately reproduces this behaviour, although
some differences are noticeable. The use of isotropic scattering, as opposed to non-isotropic scattering that tends to
be forward-throwing, may account for some (up to a factor of 2 in absolute flux) of this difference.  There are also slight
absolute offsets between the model colors and the observed colors. This offset may be due
to inaccuracies in the stellar spectrum. The differences in colors between the possible range of 
spectral types (A2 to B6, see Table \ref{parameters})
is 0.55, 0.20, 0.14 and, 0.12 magnitudes for $U-B$, $B-V$, $V-R$ and $R-I$, respectively, differences similar to or
larger than the model-data offsets. Finally, small changes in the
Johnson filter curves used in the models can produce similar offsets.

\subsection{Mid-infrared spectrum}
\label{midir}
\begin{figure}
\plotone{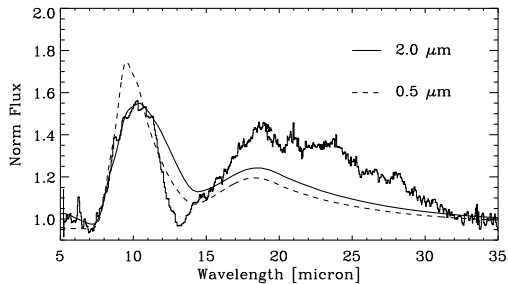}
\caption{The Spitzer spectrum of the silicate emission features of VV Ser compared to 
opacities of spherical grains with diameters of 1.0 and 3.0\,$\mu$m. A power-law continuum has been
subtracted from both the observed spectrum as well the opacity.}
\label{sed_zoom}
\end{figure}

The 5.2--37.0\,$\mu$m spectrum as observed by Spitzer-IRS is dominated by the silicate emission
bands at 9.7 and 18\,$\mu$m. 
The 9.7\,$\mu$m band is significantly broader and flatter than that of interstellar silicate grains. This is
illustrated in Fig. \ref{sed_zoom} where the shapes of the observed silicate bands are compared to those of dust opacities 
corresponding to grains of different sizes. In the figure, the disk continuum has been subracted by fitting a power law
to the spectrum at 5.5, 13 and 30-40\,$\mu$m, following the procedure of \cite{Kessler06}. 
A broadening of the 9.7\,$\mu$m silicate band is indicative of a grain 
size distribution dominated by grains larger than $\sim 1\mu$m \citep{vanBoekel05,Kessler06}.
Other emission features may appear to broaden the silicate bands. These include emission bands due to PAHs that 
in the case of VV Ser are visible at 6.2\,$\mu$m and possibly at 11.3\,$\mu$m, although this feature
may also be due to crystalline silicates  
(see also Geers et al., accepted). Because they are bright and unresolved, the PAH emission features are not related 
to the larger nebulosity discussed in Paper II and are most likely due to material associated with the disk. 
The exact dust size distribution of the ``large grain'' component 
is not strongly constrained by the silicate features, in particular not in the disk mid-plane. 
Here, we adopt a dust mixture with diameters $a_{\rm min}=0.3\,\mu$m 
and $a_{\rm max}=3.0\,\mu$m. The most important property of the large grain mixture, for the purposes of this model, is that 
it has grey optical and near-infrared opacities. As long as this condition is satisfied, the conclusions reached from 
the modeling are not sensitive to the details in the size distribution of the large grain component. It is likely that the 
dust mass of the disk has a significant population of much larger (millimeter-sized) grains \citep[e.g.][]{Dullemond05}, 
but this component is not constrained by the observations presented here. 

Another parameter constrained by the Spitzer-IRS
spectrum, central to the discussion of VV Ser as a UX Orionis class Herbig Ae star, is the slope of the mid-infrared SED. 
As suggested by \cite{Dullemond03}, if the inner-most regions of the disk are responsible
for the extinction events, the outer disk must have a scale height, $H/R$, similar to or smaller than that of the puffed-up inner rim. 
This is the definition of a self-shadowed disk, characterized by a declining
spectrum for $\lambda > 5\,\mu$m with weak silicate emission features. This type of object also corresponds
to a ``group II'' object in the classification of \cite{Meeus01}. VV Ser clearly has a mid-infrared SED
that suggests that the disk is self-shadowed. However, in the model, the declining SED is in part caused by
the low mass of the disk, i.e. to avoid predicting too much flux at $\rm \lambda > 5\,\mu m$, the models
are constrained both to be self-shadowed and to have a total gas + dust mass of $M \lesssim 10^{-5}\,M_{\odot}$. In particular,
the low mass is necessary to keep the SED declining beyond $\sim 30\,\mu$m, while the shadow of the puffed-inner rim ensures
that the SED declines between 5 and 30\,$\mu$m. 
A simple test of this was conducted by removing the puffed-up inner rim 
from the model, which resulted in a flat SED below 30\,$\mu$m. 
In the best-fitting model, 90\% (1.3 mag) of the extinction at 0.2\,$\mu$m in the disk along a sight-line inclined
at 71.5\degr  relative to the disk axis occurs within 1.5 AU. 

 Does the model suggest that the scale height of the disk has been lowered due to dust settling, or
has only the apparent dust mass been lowered? The scale height, $H_p/R$ for hydrostatic equilibrium at 50 AU is $0.09$.
This is somewhat smaller than the model scale height of $0.12$. However, this value is not 
strongly constrained, as long as the inner disk shadows the outer, it is difficult to tell the difference without
having more sensitive photometry at wavelengths longer than 70\,$\mu$m. 

\subsection{Near-infrared interferometry}
\label{interferometry}
Several interferometric measurements of the near-infrared morphology of the innermost part of the 
VV Ser system are available in the literature. \cite{Eisner03, Eisner04} presented $K$-band interferometry
observations of VV Ser performed with the Palomar testbed interferometer. These authors find a $K$-band size of
VV Ser of $\sim 3\,$mas for a generic flared disk with a puffed-up inner rim from \cite{Dullemond01}. 
This corresponds to $R_{\rm rim} \sim 0.39\times (d/260\,\mathrm{pc})\,$AU, or exactly half of the radius used in Model 1. 
In principle, this smaller radius will result in a dust temperature in the inner rim of $\sim$2000\,K, using the opacity
stipulated by the extinction events. 
While the interferometry does suffer from a poor sampling of the u-v plane, this is a discrepancy which deserves a closer
look. There are several ways of resolving the problem: the presented model includes a large cleared-out inner region
due to the high temperatures of small grains. However, the location of the rim, given a dust sublimation temperature, actually
depends on the dust opacity. A non-grey opacity will tend to
absorb more energy in the optical than it can emit in the near-infrared, increasing the temperature of the grains in the optically thin
region of the puffed-up inner rim. This will tend to erode the inner edge of the rim, as small grains are heated above their sublimation
temperature. This is reflected in the inner radius of Model 1.
Conversely, larger grains with a more grey opacity will significantly decrease the temperature at a given 
radius of the inner rim, 
and consequently $R_{\rm rim}$ will decrease for the same dust sublimation temperature. Specifically, 
using the grain opacity for the outer (>2 AU) disk rather than the grains responsible for the UX Orionis events, will
place the inner rim close to the value determined by \cite{Eisner04}, i.e. 0.3--0.4\,AU for an optically thick inner rim. 
This scenario then requires that small grains only
appear in the upper layers of the puffed-up rim, while larger grains can penetrate close to the star in a 
flatter structure. This extra copmonent of large grains within
the present inner rim does not strongly affect the model SED of VV Ser. 

\begin{figure}
\centering
  \includegraphics[width=9cm]{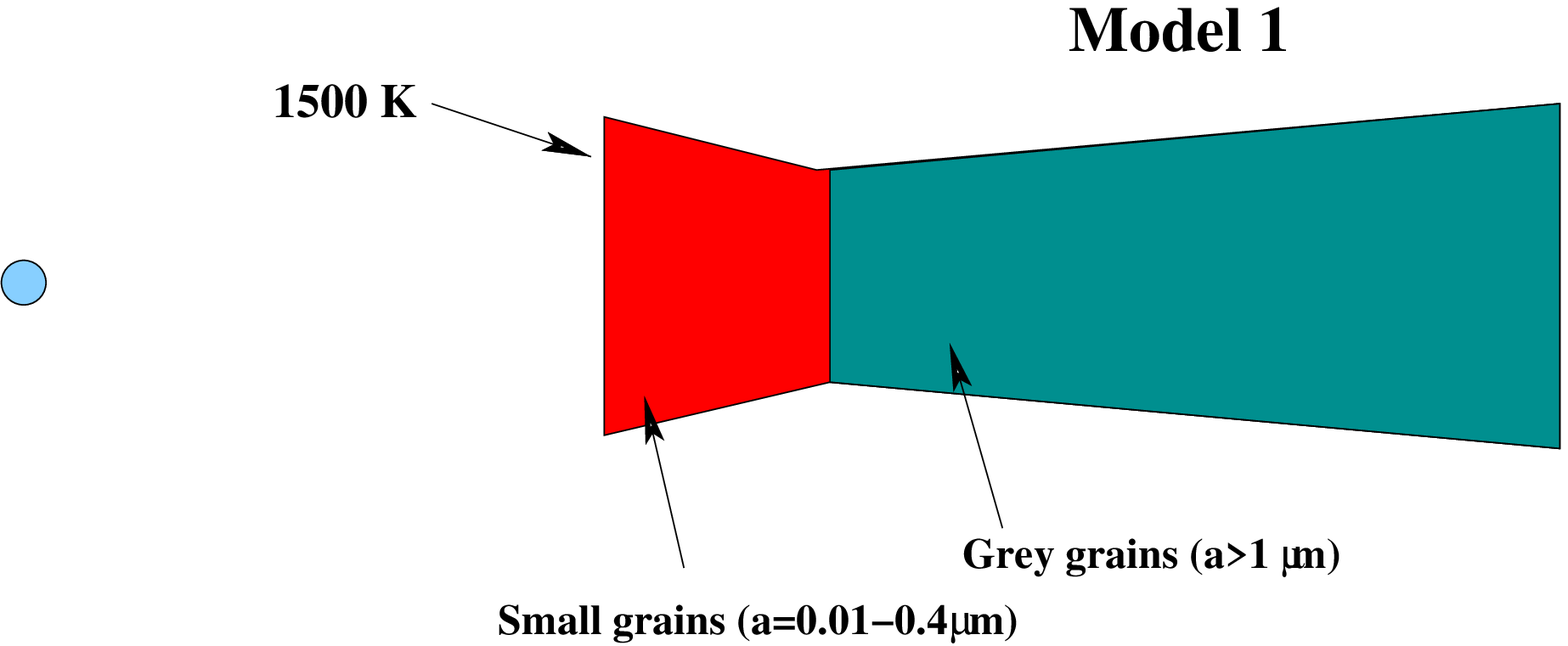}
  \includegraphics[width=9cm]{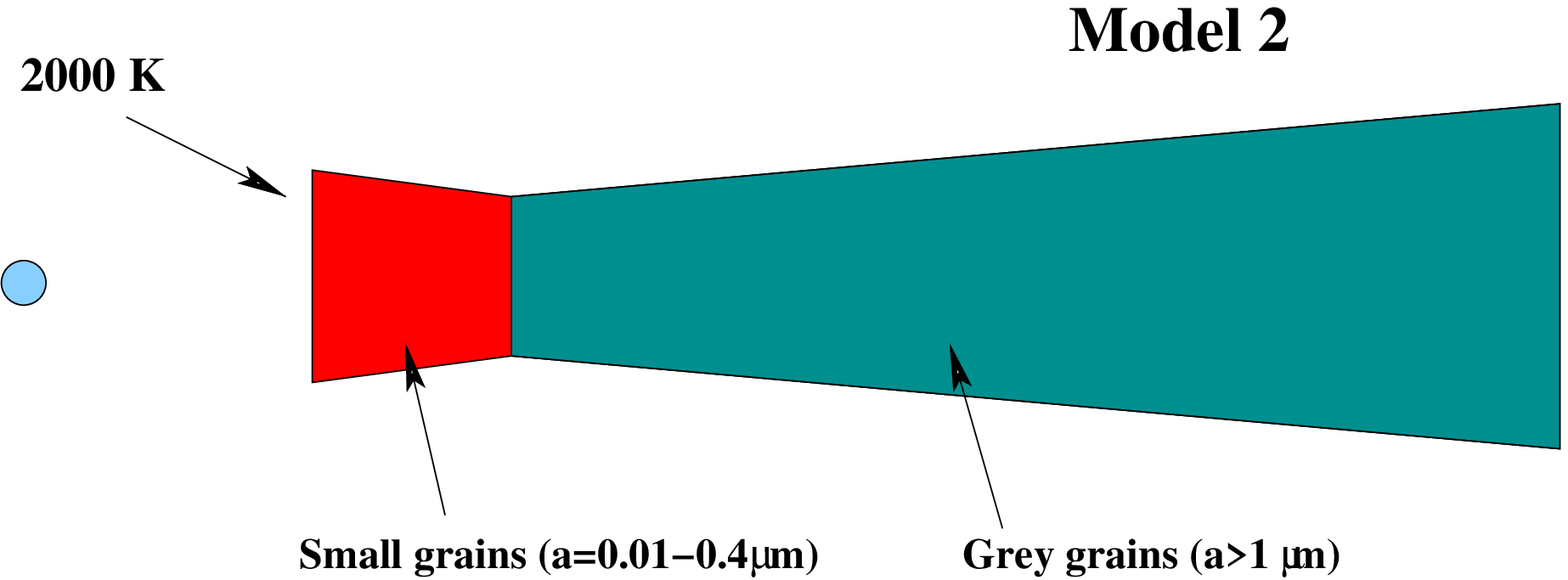}
  \includegraphics[width=9cm]{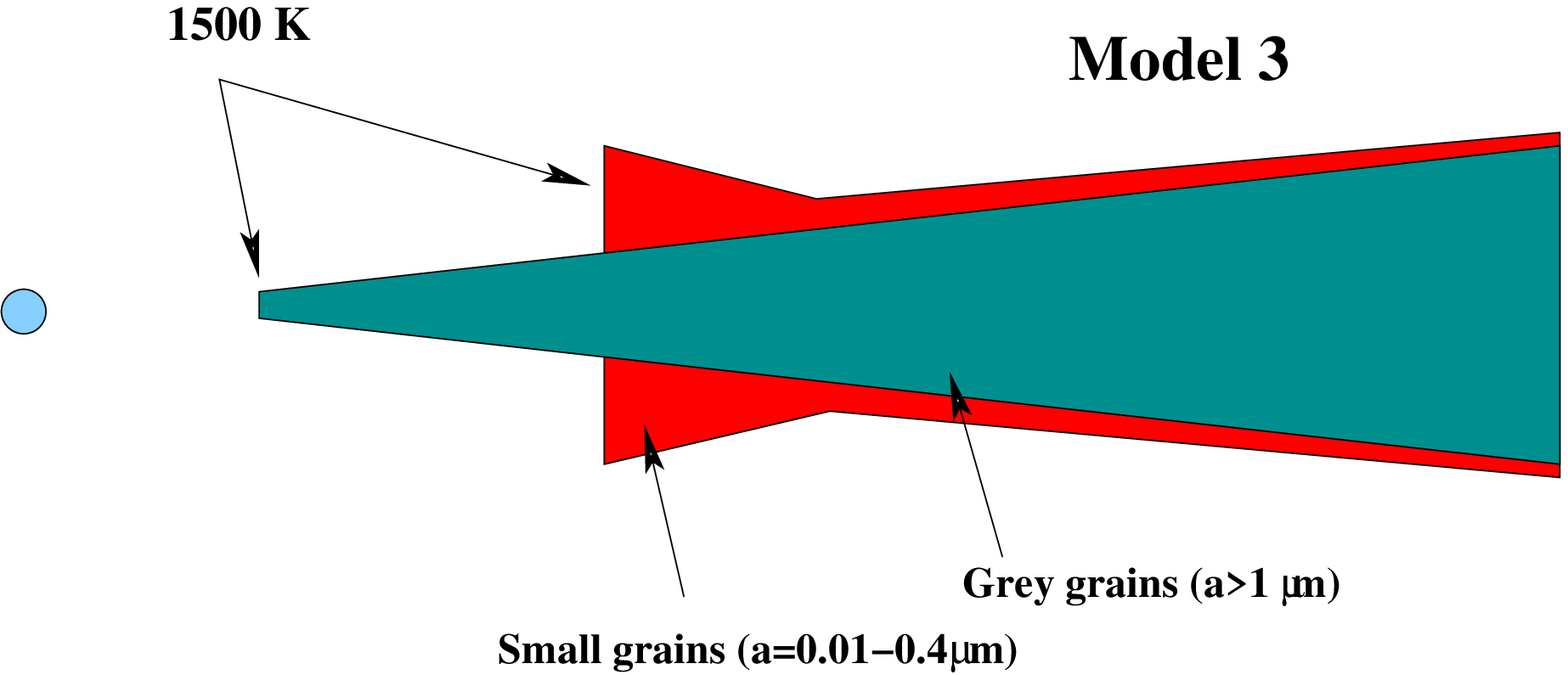}
  \caption{Sketches showing the inner rim structure for three different disk models that have been
compared to the interferometry and SED. {\it Top panel: } The ``standard'' model with the 
inner rim consisting of small grains at 1500\,K, placing it at 0.8\,AU. This model fits the SED, but not
the visibilities. {\it Middle panel: } Model using the inner rim radius of $\sim 0.4\,$AU suggested by
\cite{Eisner04}. Using the inclination and position angle from paper II as well as small grains in the rim, 
this model fits neither SED nor visibilities. The visibilities can be fit better by changing the 
position angle and inclination significantly.  {\it Bottom panel: } model using an inner rim at 1500\,K
for both grains populations present in the disk (small and grey grains). This model fits both
the SED and visibilities with the position angle and inclination from paper II.}
  \label{vis_cart}
\end{figure}

\begin{figure}
\plotone{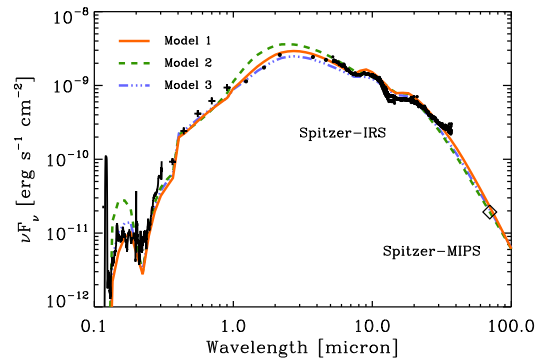}
\caption{Spectral energy distributions for VV Ser using three different models for the inner rim (see Fig. \ref{vis_cart}
and Table \ref{parameters}).
}
\label{sed_rim}
\end{figure}

\begin{figure*}
  \centering
  \includegraphics[width=12cm]{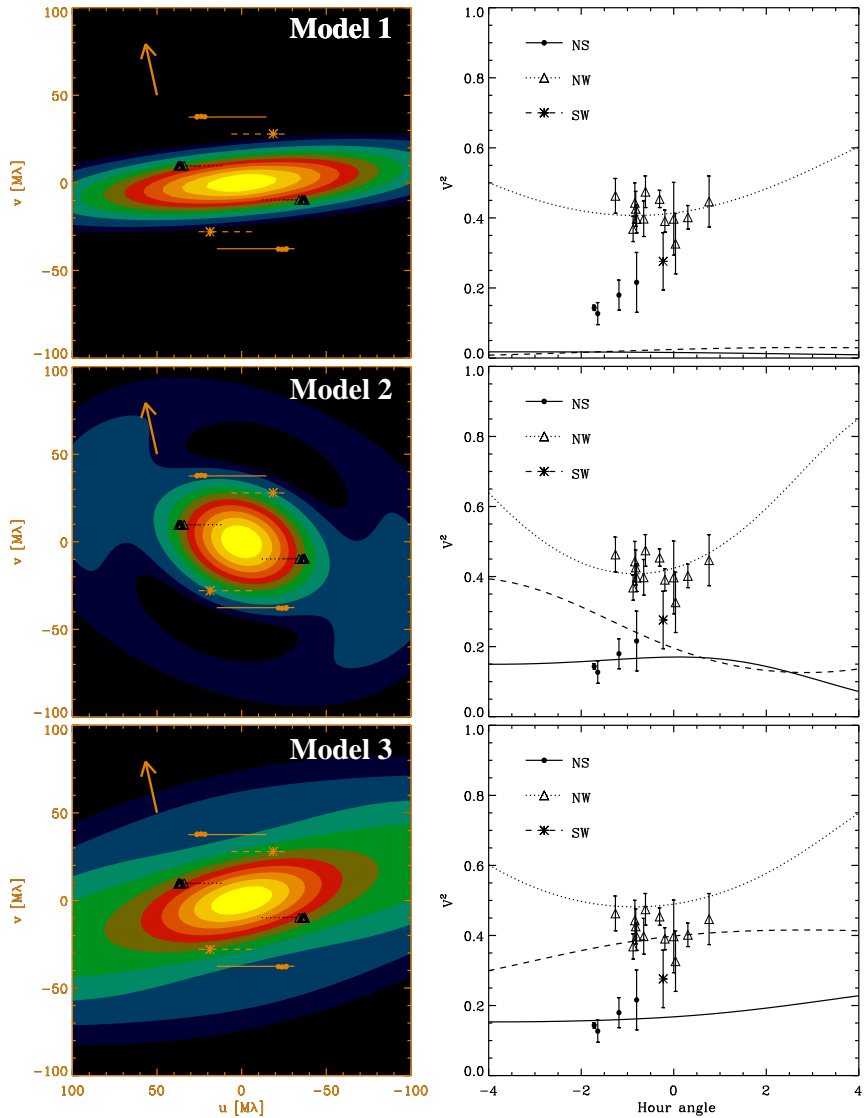}
  \caption{Normalized $K$-band visibility amplitudes of the of the ``standard'' model (Model 1), the model
with an inner rim at 0.4\,AU (Model 2) and the model simulating settling
of larger grains to the disk midplane (Model 3). The observed visibilities are from \cite{Eisner04}.
The positition angle has been varied to find the best
fit within -30 and 30\degr. The range was chosen to be reasonably consistent with the orientation of the
bipolar nebulosity around VV Ser presented in paper II.
The arrow indicates the position angle of the disk plane suggested by the {\it Spitzer} images presented in
paper II (13\degr). The symbols indicate the observed uv points for the three baselines, 
while the curves show the model tracks (left panel) and visibilities 
as a function of hour angle (right panel).}
  \label{vis}
\end{figure*}

To explore the possibilities offered by interferometry, we have calculated the visibilities for our model
as well as several alternative structures of the inner rim and compared them to those from \cite{Eisner04}.
In Fig. \ref{vis_cart}, the structures of three different types of inner rim are sketched (Model 1 refers to
the standard model discussed in the previous sections). Fig. \ref{sed_rim} shows the
calculated SEDs for the three models, while Fig. \ref{vis} shows the best fit visibilities from the three models
(allowing the position and inclination angles to vary).  
In the following, position angles are 
those of the disk plane measured east of north. It is clear
that given the position angle derived from the {\it Spitzer} images presented in paper II, the NW baseline visibilites can
be well fitted, but the NS and SW baselines produce visibilities that are severely underpredicted by Model 1, i.e. the 
visibilities are overresolved in the direction of the disk plane. 
As noted by \cite{Eisner04}, 
the single point observed with the SW baseline may be significantly more uncertain than the statistical error bar indicates.
We therefore assign little weight to this point. However, the bad fit to the NS baseline appears significant. Given the sparse uv coverage, several widely
different scenarios may provide better fits. 

First, assuming that the position angle of 10--20$\degr$ from paper II is correct, the easiest way to amend the model to fit
the near-infrared visibilities is by
making the radius of the inner rim significantly smaller, thereby increasing the visibility along the major axis of the system. 
$K$-band visibilities of a model in which the inner rim radius has artificially been
placed at 0.4\,AU has been calculated (Model 2 in Fig. \ref{vis_cart}). A significant problem with this model is that it does not fit the SED because 
the small grains in the inner rim reach temperatures of
2000\,K. Note that \cite{Eisner04} finds a temperature of 1500\,K at this radius due to an assumption of
a grey opacity for the inner rim. 
Additionally, a good fit to all three baselines can only be found for a position angle of $\sim 165\degr$ and an inclination
of 40\degr, consistent with that found by \cite{Eisner04}, but inconsistent with the orientation of the surrounding
nebulosity as described in paper II. Recently, \cite{Isella06} 
found a position angle of 60-120$\degr$ using the same interferometric data, roughly consistent with the result of \cite{Eisner04}.  
If the single uncertain point of the SW baseline is ignored, a good fit can
be found for the $\sim 70\degr$ inclination and $\sim 10\degr$ position angle found in paper II. 
These two possibilities need to be tested by further interferometric observations. 

Another option to reconcile the interferometry with the Spitzer images
is a model in which large grains in the disk mid-plane reach smaller radii in
an optically thin region within the outer puffed-up inner rim (Model 3 in Fig. \ref{vis_cart}). 
This structure is, in fact, quite plausible since the
temperature can be maintained below 1500\,K, as discussed above. This is presumably a natural scenario in a disk
in which the large grains have settled to the mid-plane.
We therefore construct a model in which the structure of the ``standard'' model is maintained (with small grains in a 
puffed-up inner rim), but with an additional component between 0.27 and 0.8\,AU consisting of the large grains of the outer disk
and with a structure following Eq. (\ref{scaleeq}). Essentially, 
this model represents a rough simulation of the effect of grain settling and density-dependent evaporation on the structure of the inner rim. 
This disk has basically two rims, both with a maximum dust temperature of 1500\,K. 
The central idea of this scenario is that the SED is not strongly affected, since both dust components have a temperature
of $\sim 1500\,$K. At the same time, the visibilities can be fitted by a disk with the same position angle and inclination
as that found in paper II using the morphology of the nebulosity seen in the IRAC and MIPS images of VV Ser. 
By tweaking additional parameters such as
the density structure of the grey dust and the exact location of the inner rim, an excellent fit to all the observables
can be obtained.
In particular, we find that a better fit to the visibilities is achieved if 
the surface density within 0.8\,AU is lowered by a factor of 5 to create an inner large grain component 
with roughly unity optical depth in the radial direction. This creates an inner rim that is not perfectly sharp, but
radiates at a range of radii. This seems consistent with the structure predicted by \cite{Isella05} using a density dependent grain
evaporation temperature.

Another option to fit the interferometer data is to add an additional point source offset by a few mas to produce additional visibility in the NW and SW baselines.
We do not attempt to model this scenario here, but simply note that a few extra visibility data points will
be able to distinguish between an additional point source and a small inner rim.

Finally, it is possible that hot gas within the inner rim may produce a significant 
contribution to the $K$-band continuum as suggested by \cite{Akeson05}, who showed models
of PTI visibilities for a small sample of T Tauri stars. In fact, gas may produce much the same
near-infrared morphology as that modeled here using two dust components. However, in order for the
gas to be sufficiently luminous, a high accretion rate of at least $10^{-7}\,M_{\odot}\,\rm yr^{-1}$ is
required \citep{Muzerolle04}. The accretion rate of VV Ser is lower than this, as evidenced by the
lack of UV excess and the very low disk mass.  

\section{Conclusions}

We have presented {\it Spitzer} mid-infrared spectroscopy of the UX Orionis star VV Ser combined with 
additional spectroscopy and imaging data from the literature spanning wavelengths from 0.1 to 850\,$\mu$m. 
Using an axisymmetric dust radiative transfer model, we have reached the following conclusions:

\begin{itemize}
\item All available data are consistent with the interpretation of the UX Orionis phenomenon in which
the central star is transiently attenuated by dust clumps in the inner puffed-up region of an inclined, self-shadowed disk
(the specific inclination is supported by the imaging data presented in Paper II). 
While VV Ser is a very specific case that has been singled out due to the availability of a wealth of 
multi-wavelength, high quality data, it provides strong support for this interpretation in the more general case. 

\item The disk appears to be both self-shadowed and low-mass, and both properties contribute to the quick decline with wavelength of the SED
at mid- and far-infrared wavelengths.
 
\item  The duration of the extinction events is consistent with an orbit
of the dust perturbations in the inner rim at a radius of $\sim 1\,$AU. 

\item The dust responsible for
the extinction events must be dominated by small, interstellar-like grains, while the dust in the outer ($>2$ AU) disk appears to be
dominated by somewhat larger grains ($\sim 1\,\mu$m). It is unlikely that the extincting dust clumps can be located significantly 
closer to the central star, since this would require them to subtend a much larger azimuthal angle
than that corresponding to the scale height of the inner rim. Additionally, they would be heated above their sublimation temperature.

\item Our best-fit inner rim radius of 0.8 AU is in contradiction with the $K$-band interferometry of \cite{Eisner04} which clearly indicates 
an inner rim radius of 0.3--0.4\,AU. A possible solution is to let the puffed-up inner rim consist
of small grains at 0.8\,AU while allowing a population consisting of larger grains to penetrate to
0.25\,AU in a mid-plane that is optically thin to infrared photons. 

\item The detailed structure of the inner rim described above is constrained {\it both} from the interferometry {\it and} the independent measurement of the disk inclination and 
position angles from large scale imaging as discussed in Paper II. 

\item We therefore interpret the 
$K$-band interferometric data of \cite{Eisner04} as evidence
for larger grains in the disk mid-plane penetrating closer to the star than the small grains in the puffed-up inner rim. 
This is consistent with a scenario in which all dust grains sublimate at $\sim 1500\,$K and larger grains have settled 
to the disk mid-plane, leaving only small grains in the disk surface. 

\item Clearly, further tests of the detailed structure of the 
puffed-up inner rim are within reach of the current observational
capabilities. Aperture synthesis observations with upcoming facilities such as VLTI-AMBER in combination with
multi-wavelength imaging and spectroscopy are certain to greatly enhance our understanding of the innermost parts
of proto-planetary disks.

\end{itemize}

\acknowledgements{The authors are grateful to Josh Eisner for providing a table with the PTI visibilities of VV Ser
and Remo Tilanus for obtaining the SCUBA map in service mode.
Support for this work was provided by NASA through Hubble Fellowship grant \#01201.01 awarded by the Space Telescope Science Institute, which is operated by the Association of 
Universities for Research in Astronomy, Inc., for NASA, under contract NAS 5-26555.
Astrochemistry in Leiden is supported by a SPINOZA grant of the Netherlands 
Organization for Scientific Research (NWO). Support for this work, part of the Spitzer Space 
Telescope Legacy Science Program, was provided by NASA through Contract Numbers 1224608 and 
1230779 issued by the Jet Propulsion Laboratory, California Institute of Technology under NASA contract 1407. 
This research was supported by the European Research Training Network ``The Origin of Planetary Systems'' 
(PLANETS, contract HPRN-CT-2002-00308). The anonymous referee is thanked for constructive comments that have significantly
improved the paper.}

\bibliographystyle{apj}
\bibliography{ms}

\begin{thebibliography}{47}
\expandafter\ifx\csname natexlab\endcsname\relax\def\natexlab#1{#1}\fi

\bibitem[{{Akeson} {et~al.}(2005){Akeson}, {Walker}, {Wood}, {Eisner}, {Scire},
  {Penprase}, {Ciardi}, {van Belle}, {Whitney}, \& {Bjorkman}}]{Akeson05}
{Akeson}, R.~L., {Walker}, C.~H., {Wood}, K., {Eisner}, J.~A., {Scire}, E.,
  {Penprase}, B., {Ciardi}, D.~R., {van Belle}, G.~T., {Whitney}, B., \&
  {Bjorkman}, J.~E. 2005, \apj, 622, 440

\bibitem[{{Berrilli} {et~al.}(1992){Berrilli}, {Corciulo}, {Ingrosso},
  {Lorenzetti}, {Nisini}, \& {Strafella}}]{Berilli92}
{Berrilli}, F., {Corciulo}, G., {Ingrosso}, G., {Lorenzetti}, D., {Nisini}, B.,
  \& {Strafella}, F. 1992, \apj, 398, 254

\bibitem[{{Bertout}(2000)}]{Bertout00}
{Bertout}, C. 2000, \aap, 363, 984

\bibitem[{{Bohren} \& {Huffman}(1983)}]{BH}
{Bohren}, C.~F., \& {Huffman}, D.~R. 1983, {Absorption and scattering of light
  by small particles} (New York: Wiley, 1983)

\bibitem[{{Bouwman} {et~al.}(2001){Bouwman}, {Meeus}, {de Koter}, {Hony},
  {Dominik}, \& {Waters}}]{Bouwman01}
{Bouwman}, J., {Meeus}, G., {de Koter}, A., {Hony}, S., {Dominik}, C., \&
  {Waters}, L.~B.~F.~M. 2001, \aap, 375, 950

\bibitem[{{Chiang} \& {Goldreich}(1997)}]{CG}
{Chiang}, E.~I., \& {Goldreich}, P. 1997, \apj, 490, 368

\bibitem[{{Draine}(2003)}]{Draine03}
{Draine}, B.~T. 2003, \araa, 41, 241

\bibitem[{{Duch{\^ e}ne} {et~al.}(2003){Duch{\^ e}ne}, {M{\' e}nard},
  {Stapelfeldt}, \& {Duvert}}]{Duchene03}
{Duch{\^ e}ne}, G., {M{\' e}nard}, F., {Stapelfeldt}, K., \& {Duvert}, G. 2003,
  \aap, 400, 559

\bibitem[{{Dullemond} \& {Dominik}(2004)}]{Dullemond04}
{Dullemond}, C.~P., \& {Dominik}, C. 2004, \aap, 417, 159

\bibitem[{{Dullemond} \& {Dominik}(2005)}]{Dullemond05}
---. 2005, \aap, 434, 971

\bibitem[{{Dullemond} {et~al.}(2001){Dullemond}, {Dominik}, \&
  {Natta}}]{Dullemond01}
{Dullemond}, C.~P., {Dominik}, C., \& {Natta}, A. 2001, \apj, 560, 957

\bibitem[{{Dullemond} \& {Turolla}(2000)}]{Dullemond00}
{Dullemond}, C.~P., \& {Turolla}, R. 2000, \aap, 360, 1187

\bibitem[{{Dullemond} {et~al.}(2003){Dullemond}, {van den Ancker}, {Acke}, \&
  {van Boekel}}]{Dullemond03}
{Dullemond}, C.~P., {van den Ancker}, M.~E., {Acke}, B., \& {van Boekel}, R.
  2003, \apjl, 594, L47

\bibitem[{{Eisner} {et~al.}(2003){Eisner}, {Lane}, {Akeson}, {Hillenbrand}, \&
  {Sargent}}]{Eisner03}
{Eisner}, J.~A., {Lane}, B.~F., {Akeson}, R.~L., {Hillenbrand}, L.~A., \&
  {Sargent}, A.~I. 2003, \apj, 588, 360

\bibitem[{{Eisner} {et~al.}(2004){Eisner}, {Lane}, {Hillenbrand}, {Akeson}, \&
  {Sargent}}]{Eisner04}
{Eisner}, J.~A., {Lane}, B.~F., {Hillenbrand}, L.~A., {Akeson}, R.~L., \&
  {Sargent}, A.~I. 2004, \apj, 613, 1049

\bibitem[{{Evans} {et~al.}(2003){Evans}, {Allen}, {Blake}, {Boogert}, {Bourke},
  {Harvey}, {Kessler}, {Koerner}, {Lee}, {Mundy}, {Myers}, {Padgett},
  {Pontoppidan}, {Sargent}, {Stapelfeldt}, {van Dishoeck}, {Young}, \&
  {Young}}]{Evans03}
{Evans}, N.~J., {Allen}, L.~E., {Blake}, G.~A., {Boogert}, A.~C.~A., {Bourke},
  T., {Harvey}, P.~M., {Kessler}, J.~E., {Koerner}, D.~W., {Lee}, C.~W.,
  {Mundy}, L.~G., {Myers}, P.~C., {Padgett}, D.~L., {Pontoppidan}, K.,
  {Sargent}, A.~I., {Stapelfeldt}, K.~R., {van Dishoeck}, E.~F., {Young},
  C.~H., \& {Young}, K.~E. 2003, \pasp, 115, 965

\bibitem[{{Grady} {et~al.}(1997){Grady}, {Sitko}, {Bjorkman}, {Perez}, {Lynch},
  {Russell}, \& {Hanner}}]{Grady97}
{Grady}, C.~A., {Sitko}, M.~L., {Bjorkman}, K.~S., {Perez}, M.~R., {Lynch},
  D.~K., {Russell}, R.~W., \& {Hanner}, M.~S. 1997, \apj, 483, 449

\bibitem[{{Grinin}(1988)}]{Grinin88}
{Grinin}, V.~P. 1988, Pis ma Astronomicheskii Zhurnal, 14, 65

\bibitem[{{Hern{\' a}ndez} {et~al.}(2004){Hern{\' a}ndez}, {Calvet}, {Brice{\~
  n}o}, {Hartmann}, \& {Berlind}}]{Hernandez04}
{Hern{\' a}ndez}, J., {Calvet}, N., {Brice{\~ n}o}, C., {Hartmann}, L., \&
  {Berlind}, P. 2004, \aj, 127, 1682

\bibitem[{{Houck} {et~al.}(2004){Houck}, {Roellig}, {van Cleve}, {Forrest},
  {Herter}, {Lawrence}, {Matthews}, {Reitsema}, {Soifer}, {Watson}, {Weedman},
  {Huisjen}, {Troeltzsch}, {Barry}, {Bernard-Salas}, {Blacken}, {Brandl},
  {Charmandaris}, {Devost}, {Gull}, {Hall}, {Henderson}, {Higdon}, {Pirger},
  {Schoenwald}, {Sloan}, {Uchida}, {Appleton}, {Armus}, {Burgdorf},
  {Fajardo-Acosta}, {Grillmair}, {Ingalls}, {Morris}, \& {Teplitz}}]{Houck04}
{Houck}, J.~R., {Roellig}, T.~L., {van Cleve}, J., {Forrest}, W.~J., {Herter},
  T., {Lawrence}, C.~R., {Matthews}, K., {Reitsema}, H.~J., {Soifer}, B.~T.,
  {Watson}, D.~M., {Weedman}, D., {Huisjen}, M., {Troeltzsch}, J., {Barry},
  D.~J., {Bernard-Salas}, J., {Blacken}, C.~E., {Brandl}, B.~R.,
  {Charmandaris}, V., {Devost}, D., {Gull}, G.~E., {Hall}, P., {Henderson},
  C.~P., {Higdon}, S.~J.~U., {Pirger}, B.~E., {Schoenwald}, J., {Sloan}, G.~C.,
  {Uchida}, K.~I., {Appleton}, P.~N., {Armus}, L., {Burgdorf}, M.~J.,
  {Fajardo-Acosta}, S.~B., {Grillmair}, C.~J., {Ingalls}, J.~G., {Morris},
  P.~W., \& {Teplitz}, H.~I. 2004, \apjs, 154, 18

\bibitem[{{Isella} \& {Natta}(2005)}]{Isella05}
{Isella}, A., \& {Natta}, A. 2005, \aap, 438, 899

\bibitem[{{Isella} {et~al.}(2006){Isella}, {Testi}, \& {Natta}}]{Isella06}
{Isella}, A., {Testi}, L., \& {Natta}, A. 2006, \aap, 451, 951

\bibitem[{{J{\"a}ger} {et~al.}(1998){J{\"a}ger}, {Mutschke}, \&
  {Henning}}]{Jaeger98}
{J{\"a}ger}, C., {Mutschke}, H., \& {Henning}, T. 1998, \aap, 332, 291

\bibitem[{{Kenyon} {et~al.}(1993){Kenyon}, {Calvet}, \& {Hartmann}}]{Kenyon93}
{Kenyon}, S.~J., {Calvet}, N., \& {Hartmann}, L. 1993, \apj, 414, 676

\bibitem[{{Kessler-Silacci} {et~al.}(2006){Kessler-Silacci}, {Augereau},
  {Dullemond}, {Geers}, {Lahuis}, {Evans}, {van Dishoeck}, {Blake}, {Boogert},
  {Brown}, {J{\o}rgensen}, {Knez}, \& {Pontoppidan}}]{Kessler06}
{Kessler-Silacci}, J., {Augereau}, J.-C., {Dullemond}, C.~P., {Geers}, V.,
  {Lahuis}, F., {Evans}, II, N.~J., {van Dishoeck}, E.~F., {Blake}, G.~A.,
  {Boogert}, A.~C.~A., {Brown}, J., {J{\o}rgensen}, J.~K., {Knez}, C., \&
  {Pontoppidan}, K.~M. 2006, \apj, 639, 275

\bibitem[{{Meeus} {et~al.}(2001){Meeus}, {Waters}, {Bouwman}, {van den Ancker},
  {Waelkens}, \& {Malfait}}]{Meeus01}
{Meeus}, G., {Waters}, L.~B.~F.~M., {Bouwman}, J., {van den Ancker}, M.~E.,
  {Waelkens}, C., \& {Malfait}, K. 2001, \aap, 365, 476

\bibitem[{{Meijerink} {et~al.}(2005){Meijerink}, {Tilanus}, {Dullemond},
  {Israel}, \& {van der Werf}}]{Meijerink05}
{Meijerink}, R., {Tilanus}, R.~P.~J., {Dullemond}, C.~P., {Israel}, F.~P., \&
  {van der Werf}, P.~P. 2005, \aap, 430, 427

\bibitem[{{Mora} {et~al.}(2001){Mora}, {Mer{\'{\i}}n}, {Solano}, {Montesinos},
  {de Winter}, {Eiroa}, {Ferlet}, {Grady}, {Davies}, {Miranda}, {Oudmaijer},
  {Palacios}, {Quirrenbach}, {Harris}, {Rauer}, {Cameron}, {Deeg}, {Garz{\'
  o}n}, {Penny}, {Schneider}, {Tsapras}, \& {Wesselius}}]{Mora01}
{Mora}, A., {Mer{\'{\i}}n}, B., {Solano}, E., {Montesinos}, B., {de Winter},
  D., {Eiroa}, C., {Ferlet}, R., {Grady}, C.~A., {Davies}, J.~K., {Miranda},
  L.~F., {Oudmaijer}, R.~D., {Palacios}, J., {Quirrenbach}, A., {Harris},
  A.~W., {Rauer}, H., {Cameron}, A., {Deeg}, H.~J., {Garz{\' o}n}, F., {Penny},
  A., {Schneider}, J., {Tsapras}, Y., \& {Wesselius}, P.~R. 2001, \aap, 378,
  116

\bibitem[{{Muzerolle} {et~al.}(2004){Muzerolle}, {D'Alessio}, {Calvet}, \&
  {Hartmann}}]{Muzerolle04}
{Muzerolle}, J., {D'Alessio}, P., {Calvet}, N., \& {Hartmann}, L. 2004, \apj,
  617, 406

\bibitem[{{Natta} {et~al.}(2001){Natta}, {Prusti}, {Neri}, {Wooden}, {Grinin},
  \& {Mannings}}]{Natta01}
{Natta}, A., {Prusti}, T., {Neri}, R., {Wooden}, D., {Grinin}, V.~P., \&
  {Mannings}, V. 2001, \aap, 371, 186

\bibitem[{{Natta} \& {Whitney}(2000)}]{Natta00}
{Natta}, A., \& {Whitney}, B.~A. 2000, \aap, 364, 633

\bibitem[{{Ossenkopf} \& {Henning}(1994)}]{OH94}
{Ossenkopf}, V., \& {Henning}, T. 1994, \aap, 291, 943

\bibitem[{{Pascucci} {et~al.}(2004){Pascucci}, {Wolf}, {Steinacker},
  {Dullemond}, {Henning}, {Niccolini}, {Woitke}, \& {Lopez}}]{Pascucci04}
{Pascucci}, I., {Wolf}, S., {Steinacker}, J., {Dullemond}, C.~P., {Henning},
  T., {Niccolini}, G., {Woitke}, P., \& {Lopez}, B. 2004, \aap, 417, 793

\bibitem[{Pontoppidan \& Dullemond(2005)}]{Pontoppidan_shadow}
Pontoppidan, K.~M., \& Dullemond, C.~P. 2005, \aap, 435, 611

\bibitem[{Pontoppidan {et~al.}(2006)Pontoppidan, Dullemond, \&
  et~al.}]{paperII}
Pontoppidan, K.~M., Dullemond, C.~P., \& et~al. 2006, \apj

\bibitem[{Pontoppidan {et~al.}(2005)Pontoppidan, Dullemond, van Dishoeck,
  Blake, Boogert, Evans~II, Kessler-Silacci, \& Lahuis}]{Pontoppidan_crbr}
Pontoppidan, K.~M., Dullemond, C.~P., van Dishoeck, E.~F., Blake, G.~A.,
  Boogert, A. C.~A., Evans~II, N.~J., Kessler-Silacci, J.~E., \& Lahuis, F.
  2005, \apj, 622, 463

\bibitem[{{Rostopchina} \& {Grinin}(2001)}]{Rostopchina01}
{Rostopchina}, A.~N., \& {Grinin}, V.~P. 2001, Astronomy Reports, 45, 51

\bibitem[{{Siess} {et~al.}(2000){Siess}, {Dufour}, \& {Forestini}}]{Siess00}
{Siess}, L., {Dufour}, E., \& {Forestini}, M. 2000, \aap, 358, 593

\bibitem[{{Straizys} {et~al.}(1996){Straizys}, {Cernis}, \&
  {Bartasiute}}]{Straizys96}
{Straizys}, V., {Cernis}, K., \& {Bartasiute}, S. 1996, Baltic Astronomy, 5,
  125

\bibitem[{{Thi} {et~al.}(2002){Thi}, {Pontoppidan}, {van Dishoeck}, {Dartois},
  \& {d'Hendecourt}}]{Thi02}
{Thi}, W.~F., {Pontoppidan}, K.~M., {van Dishoeck}, E.~F., {Dartois}, E., \&
  {d'Hendecourt}, L. 2002, \aap, 394, L27

\bibitem[{{van Bemmel} \& {Dullemond}(2003)}]{vanBemmel03}
{van Bemmel}, I.~M., \& {Dullemond}, C.~P. 2003, \aap, 404, 1

\bibitem[{{van Boekel} {et~al.}(2005){van Boekel}, {Min}, {Waters}, {de Koter},
  {Dominik}, {van den Ancker}, \& {Bouwman}}]{vanBoekel05}
{van Boekel}, R., {Min}, M., {Waters}, L.~B.~F.~M., {de Koter}, A., {Dominik},
  C., {van den Ancker}, M.~E., \& {Bouwman}, J. 2005, \aap, 437, 189

\bibitem[{{van den Ancker}(1999)}]{vandenancker99}
{van den Ancker}, M. 1999, Ph.D.~Thesis

\bibitem[{{Watson} \& {Stapelfeldt}(2004)}]{Watson04}
{Watson}, A.~M., \& {Stapelfeldt}, K.~R. 2004, \apj, 602, 860

\bibitem[{{Werner} {et~al.}(2004){Werner}, {Roellig}, {Low}, {Rieke}, {Rieke},
  {Hoffmann}, {Young}, {Houck}, {Brandl}, {Fazio}, {Hora}, {Gehrz}, {Helou},
  {Soifer}, {Stauffer}, {Keene}, {Eisenhardt}, {Gallagher}, {Gautier}, {Irace},
  {Lawrence}, {Simmons}, {Van Cleve}, {Jura}, {Wright}, \&
  {Cruikshank}}]{Werner04}
{Werner}, M.~W., {Roellig}, T.~L., {Low}, F.~J., {Rieke}, G.~H., {Rieke}, M.,
  {Hoffmann}, W.~F., {Young}, E., {Houck}, J.~R., {Brandl}, B., {Fazio}, G.~G.,
  {Hora}, J.~L., {Gehrz}, R.~D., {Helou}, G., {Soifer}, B.~T., {Stauffer}, J.,
  {Keene}, J., {Eisenhardt}, P., {Gallagher}, D., {Gautier}, T.~N., {Irace},
  W., {Lawrence}, C.~R., {Simmons}, L., {Van Cleve}, J.~E., {Jura}, M.,
  {Wright}, E.~L., \& {Cruikshank}, D.~P. 2004, \apjs, 154, 1

\bibitem[{{Wolf} {et~al.}(2003){Wolf}, {Padgett}, \& {Stapelfeldt}}]{Wolf03}
{Wolf}, S., {Padgett}, D.~L., \& {Stapelfeldt}, K.~R. 2003, \apj, 588, 373

\bibitem[{{Wood} {et~al.}(2002){Wood}, {Wolff}, {Bjorkman}, \&
  {Whitney}}]{Wood02}
{Wood}, K., {Wolff}, M.~J., {Bjorkman}, J.~E., \& {Whitney}, B. 2002, \apj,
  564, 887

\end{thebibliography}

\end{document}